\documentclass[12pt]{article}
\usepackage{amsmath}
\usepackage{amsfonts}
\usepackage{epsfig}
\renewcommand{\phi}{\varphi}

\newcommand{\beq}{\begin{equation}}
\newcommand{\eeq}{\end{equation}}

\newcommand{\pd}{\partial}

\newcommand{\bphi}{\bar{\phi}}

\newcommand{\bg}{\bar{g}}
\newcommand{\bdel}{\bar{\delta}}
\newcommand{\balpha}{\bar{\alpha}}

\newcommand{\ep}{\epsilon}
\newcommand{\lra}{\leftrightarrow}

\newcommand{\cotan}{{\rm cotan}}

\title{\bf The Regge-Gribov model with odderons.}
\author{M.A. Braun$^1$, E.M. Kuzminskii$^{2}$, M.I. Vyazovsky$^1$\\
$^1${\small Dept. of High Energy Physics, Saint-Petersburg State University,
St. Petersburg, Russia}\\
$^2${\small Theoretical Physics Division,
Petersburg Nuclear Physics Institute, Gatchina, Russia}}

\begin{document}
\maketitle
{\bf Abstract}
The Regge-Gribov model describing interacting pomerons and odderons is proposed with triple reggeon vertices
taking into account the negative signature of the odderon. Its simplified version with zero transverse dimensions is first considered.
No phase transition occurs in this case at the  intercept crossing unity. This simplified model is studied
without more approximations by numerical techniques.
The physically relevant model in the two-dimensional transverse space is then
studied by the renormalization group method in the single loop approximation.
The pomeron and odderon are taken to have different bare intercepts and slopes.
The behaviour when the intercepts move from below to their
critical values compatible with the Froissart limitation is studied.
Five real fixed points are found with  singularities in the form of non-trivial branch points
indicating a phase transition as the intercepts cross unity.
The new phases, however, are not physical, since they violate the projectile-target symmetry.
In the vicinity of fixed points the asymptotical behaviour of Green functions and elastic scattering amplitude is found
under Glauber approximation for couplings to participants.

\noindent
\noindent

\section{Introduction}
In the kinematic region
where the energy is much greater than transferred momenta (''the Regge
kinematics'')  strong interactions can be phenomenologically
described by the exchange of reggeons, which correspond to poles in the
complex angular momentum plane. In this framework the high-energy asymptotic
is governed by the exchange of pomerons whose interaction can be described
by the theory introduced by V.N.Gribov with the triple pomeron vertex and an
imaginary coupling constant. Much attention was given to the study of this
theory in the past \cite{gribov,migdal,migdal1}
and more recently in \cite{bartels,vacca}.
Long ago this theory was also   applied to the study of
 $pA$ interaction at high energies in ~\cite{schwimmer}, where the sum of all
fan diagrams  was found (similar to the later treatment in the QCD framework,
which lead to the well-known Balitski-Kovchegov equations ~\cite{bal,kov,kov1}).

The reggeon theory is, however, still  a
full-fletched quantum field theory and does not allow to find  scattering amplitudes constructively.
Because of this a  simpler model in the zero-dimensional transverse world
(''toy'' model) was considered and studied in some detail
\cite{amati1,amati2,jengo,amati3,rossi,braun1,bondarenko,braun3,braun2}.
 The important messages which
followed from these studies were that 1) the quantum effects, that is
loops, change cardinally the high-energy behaviour of the amplitudes
and so their neglect is  a very crude approximation
and 2) passage through the intercept $\alpha_P(0)=1$ goes
smoothly, without phase transition, so that the theory preserves its physical sense
for the supercritical pomeron with $\alpha_P(0)>1$.

The second of these important findings has been, however, found  wrong in the physical
case of two transverse dimensions. Using the renormalization group (RG) technique in
\cite{abarb2} it was concluded that at $\alpha_P(0)=1$ a second order phase transition occurs.
New phases, which arise at $\alpha_P(0)>1$, cannot be considered  physical, since they violate
the fundamental symmetry target-projectile. So the net result was that the model
could not accommodate the supercritical pomeron with $\alpha_P(0)>1$ altogether.

In the QCD, apart from the pomeron with the
positive $C$-parity and signature, a compound state of three reggeized gluons with
the negative $C$-parity and signature, the odderon, appears. Actually it was proposed before the QCD era
on general grounds in ~\cite{nicol}. Since then its  possible experimental
manifestations have been widely discussed
~\cite{odderon1,odderon2,odderon3} with  conclusions containing a large dose of uncertainty up to now, which may be
explained both by the difficulties in the experimental settings and  the elusive properties of the odderon itself.
On the theoretical level the QCD  odderon was discussed in many papers \cite{jw1,jw2,blv,hiim}.
Its intercept was found to lie in the vicinity of unity  and  the  coupling constants for its interactions were
guessed to be the same as for the pomeron interactions.

In this study we introduce the Regge-Gribov model containing odderons apart from pomerons
with the interaction between them following the rule of signature conservation.
We first consider the simplified picture in the zero-dimensioned transverse space.
Passing to the physical two-dimensional transverse space we apply the renormalization group techniques to study asymptotical
properties of the model in the vicinity of fixed points in the single loop approximation.
Our results were partially published in ~\cite{bkv2023,bkv2025}.

\section{Model}
Our model describes two complex fields $\phi_{1,2}$ for the pomeron $\phi_1$ and odderon $\phi_2$ acting in the D-dimensional transverse space
and depending on rapidity $y$
with the Lagrangian
\[
{\cal L}=\sum_{i=1}^2\Big(\bphi_{i}\pd_y\phi_{i}-\mu_{i}\bphi_{i}\phi_{i}+\alpha'_{i}\nabla\bphi_{i}\nabla\phi_{i}\Big)\]\beq
+\frac{i}{2}\Big(\lambda_{1}\bphi_{1}(\phi_{1}+\bphi_{1})\phi_{1}+2\lambda_{2}(\bphi_{2}\phi_{2}(\bphi_{1}+\phi_{1}))
+\lambda_{3}(\bphi_{2}^2\phi_{1}-\phi_{2}^2\bphi_{1})\Big).
\label{eq1}
\eeq
It contains two different bare "masses" $\mu_{1}$ and $\mu_{2}$ and slope parameters $\alpha'_{i}$ for the pomeron and odderon.
The masses are defined as the intercepts minus unity. In the free theory with $\lambda_i=0$
one has $\alpha_i(0)=1+\mu_{i}$, $i=1,2$.
This  Lagrangian is constructed in accordance with signature conservation rule.
One can ascribe parity $P=+1$ and  $P=-1$ for the pomeron and odderon fields, respectively.
The Lagrangian turns out to be parity invariant, so that conservation of signature becomes translated into conservation of parity.
The Lagrangian is also  invariant under transformation
\beq \phi_1(y,x)\lra \bphi_1(-y,x),\ \ \phi_2(y,x)\lra i\bphi_2(-y,x),
\label{invj4}
\eeq
which reflects the symmetry between the projectile and target.
It is supplemented by the external coupling to participants in the form
\[
{\cal L}_{ext}(y,x)=\delta(y-Y)\Big(i\rho_p^{(P)}(x)\phi_1(y,x)+\rho_p^{(O)}(x)\phi_2(y,x\Big)\]\beq+
\delta(y)\Big(i\rho^{(P)}_t(x)\bphi_1(y,x))+i\rho_t^{(O)}(x)\bphi_2(y,x)\Big),
\label{lext}
\eeq
where it is assumed that the projectile has rapidity $Y$ and the target is at rest.

Note that in our model in the lowest approximation, when one neglects all interactions between pomerons and odderons,
the single reggeon exchange is reduced just to their propagators. Actually, however, this contribution  is given by some
external couplings squared
multiplied by appropriate signature factors which are (up to unimportant coefficients)
\beq
\zeta_P
=-i-\tan (\pi\mu_1/2),\ \ \zeta_O
=-\cotan(\pi\mu_2/2)+i\label{zeta}\eeq
for the pomeron and  odderon, respectively.
Both are complex and have real and imaginary parts. In our picture these factors have to be included into the
external couplings, so that the coupling of a single reggeon to the participants should be in fact
$g_P^2=ig_{P0}^2\zeta_P,\ \ g_O^2=ig_{O0}^2\zeta_O$
where $g_{P0}$ and $g_{O0}$ are real constants corresponding to the couplings in the original Regge-Gribov picture.
Note that at small masses  the pomeron coupling turns out to be real positive whereas it is positive imaginary for  the odderon,
in agreement with their contributions in the standard approach. With interactions turned on  the couplings are assumed
to stay intact. However, the masses run and bare masses which figure in (\ref{zeta}) become infinite.
So we cannot naively include signature factors into the coupling constants. Instead we assume some fixed values for $g_P^2$ an $g_O^2$
admitting that they can be complex.

The  amplitude ${\cal A}$  is standardly given by
\beq
{\cal A}(Y)=-i\Big<{\rm T}\Big\{e^{\int dyd^Dx{\cal L}_{ext}} S_{int}\Big\}\Big> ,
\label{ampli}
\eeq
where $S_{int}$ is the  $S$ matrix in the interaction representation.
Alternatively, the amplitude can be expressed via the corresponding Hamiltonian as
\beq
i{\cal A}(Y)=<\Psi_{fin}|e^{-HY}|\Psi_{in}>,\label{ampli1}\eeq
where initial and final states are given by
\beq
\Psi_{fin}=1-\exp \Big[\int d^Dx\Big(i\rho^{(P)}_p(x)\phi_1(Y,x)+\rho_p^{(O)}(x)\phi_2(Y,x\Big)\Big]
\eeq
and
\beq
\Psi_{in}=1-\exp \Big[\int d^Dx\Big(i\rho^{(P)}_t(x)\bphi_1(0,x)+i\rho_t^{(O)}(x)\bphi_2(0,x\Big)\Big].
\eeq

The ground state of the Hamiltonian derived from (\ref{eq1}) takes different forms depending on the existence of the vacuum averages
of the pomeron field. As shown in ~\cite{abarb2} in absence of odderons three of them lead to kinematical terms admitting
 valid representation via functional integral and perturbative treatment.
In absence of the vacuum average the pomeron mass term  is    $\mu\bphi_1\phi_1$ and the theory is good at $\mu\leq 0$.  The two other ground states
correspond to $<\phi_1>\neq 0$ but $<\bphi_1>=0$ or vice versa. In both cases the pomeron mass term aquires  the minus sign to be
$-\mu\bphi_1\phi_1$.
So these phases admit the perturbative treatment for $\mu\geq 0$. However, these phases violate the projectile-target symmetry and so are unphysical.
Inclusion of odderons does not change these conclusions, since by parity conservation the odderon field cannot possess a non-zero vacuum average.
So in the following phases with nonzero vacuum averages  will not be considered and we start with $<\phi_i>=<\bphi_i>=0$, $i=1,2$ and $\mu\leq 0$
attempting to analytically continue the theory for positive $\mu$.

Treated perturbatively,  the theory contains divergent diagrams for self-masses and triple vertices and so needs renormalization.
This will be studied in Section 4. Meanwhile we shall consider the simplified model with zero transverse dimensions $D=0$,
which does not contain divergences.

\section{Zero transverse dimensions}
With zero transverse dimensions the fields become functions of only rapidity and the model essentially transforms into the quantum-mechanical one.
For still more simplification we assume the three coupling constants in (\ref{eq1}) equal, to be $2\lambda>0$, which corresponds to conclusions from the QCD.
To pass to real variables it is convenient to introduce new operators $u$, $v$, $w$ and $z$ putting
\[ \phi_1=-iv,\ \ \phi_1^\dagger=-iu,\ \ \phi_2=-iz,\ \ \phi_2^\dagger=-iw\]
with  abnormal commutation relations
\[[v,u]=[z,w]=-1.\]
In terms of these  variables the Hamiltonian becomes real
\beq
H=
\mu_1uv+\mu_2wz-
\lambda\Big(u^2v+uv^2+2wuz+2wzv-uz^2+w^2v\Big)\label{hpo}\eeq
with the vacuum state satisfying
\[v|0>=z|0>=0.\]
We put the odderon mass $\mu_2=0$, again following the QCD predictions.

Without odderons this theory was thoroughly studied long ago. It was found that by a certain change of variables the Hamiltonian could be transformed to
the hermitian form~\cite{jengo} and that the theory admits a smooth continuation from $\mu<0$ to $\mu>0$ (see e.g. ~\cite{amati3}). For $\mu>0$ and small $\lambda>0$
the ground state energy was found to be positive and equal to
\beq
E_P=\frac{\mu\rho}{\sqrt{2\pi}}e^{-\rho^2/2}\Big(1-\frac{2}{\rho^2}+{\cal O}(\frac{1}{\rho^4})\Big)
\label{ep}
\eeq
where $\rho=\mu/\lambda$.
More energy levels were found in ~\cite{braun3} from the asymptotical behaviour of the Schroedinger equation in original variables
(equivalent to the biconfluent Heun equation).
With the known  energy levels  the behaviour of the amplitudes in rapidity can be standardly obtained by developing in the eigenstates of the
Hamiltonian.

An alternative procedure was proposed in ~\cite{braun1}. In the $u$ representation $v=-\pd/\pd u$, so that the
Schroedinger equation converts to the evolution equation in rapidity
\beq
\frac{\pd \Psi(y,u)}{\pd y}=-H\Psi(y,u)\label{evol}
\eeq
where $H$ is the second order differential operator in $u$. Taking  the initial function $\Psi(0,u)$ and using (\ref{evol})
one can find $\Psi(Y,u)$ numerically at any $Y$. For practical applications  this method is simple and productive  although it
is less suitable to study far asymptotics, where analytic methods could be preferable.

Inclusion of odderons spoils this simple picture, since one cannot find adequate change of variables to reduce the problem
to the Schroedinger equation with a hermitian Hamiltonian. So the only possibility is to apply numerical evolution of the wave function
$\Psi(y,u,w)$  using Hamiltonian (\ref{hpo}). However, at this point we meet with the difficulty. Unlike the pure pomeron case, we do not know the domain
in the $u,w$ plane where the wave function is defined. Attempts to take both $u$ and $w$ real positive fail, which manifests itself by a
complete breakdown of evolution at already quite small rapidities. It turns out~\cite{bkv} that transformation to new fields $\phi$ and $\psi$
\[\phi=\frac{\phi_1+i\phi_2}{\sqrt{2}},\ \ \psi=\frac{\phi_1-i\phi_2}{\sqrt{2}},\ \ \phi^\dagger=\frac{\phi_1^\dagger-i\phi_2^\dagger}{\sqrt{2}},\ \
\psi^\dagger=\frac{\phi_1^\dagger+i\phi_2^\dagger}{\sqrt{2}}\]
solves this problem with the new variables  both real and positive. Evolution then safely works up to quite high values of rapidity.
To illustrate our results, in Fig. 1 we present the pomeron and odderon propagators for $\mu_1=1$ and $\lambda=1/3$.
As one observes the influence of the odderon is to enhance the pomeron propagator at high rapidities.
\begin{figure}[h!p]
\begin{center}
\includegraphics[width=7.2 cm]{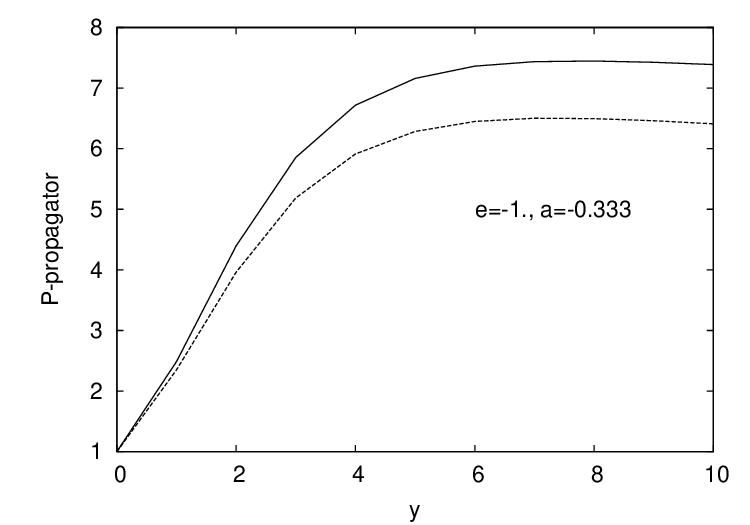}
\includegraphics[width=7.2 cm]{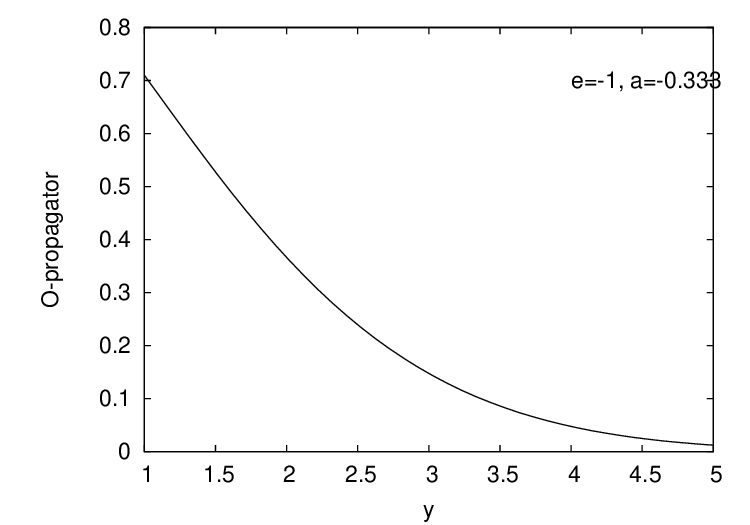}
\caption{The solid curves show pomeron (left panel) and odderon (right panel) propagators
as functions of rapidity for $\mu_1=1$, $\lambda=1/3$. The dashed curve in the left panel
shows the pomeron propagator in absence of the odderon (only pomeron loops).}
\label{fig1}
\end{center}
\end{figure}

It is of interest to know the behaviour at very high energies. To this aim we calculated the ground state energy
in the model with odderons from the behaviour of the pomeron propagator at rapidities up to $y=50$ obtained numerically.
We checked that this value is stable within the evolution range 30$\div$50.
The found ratio $E/E_P$ for different $\rho$ is shown in Fig. 2.
\begin{figure}[h]
\begin{center}
\epsfig{file=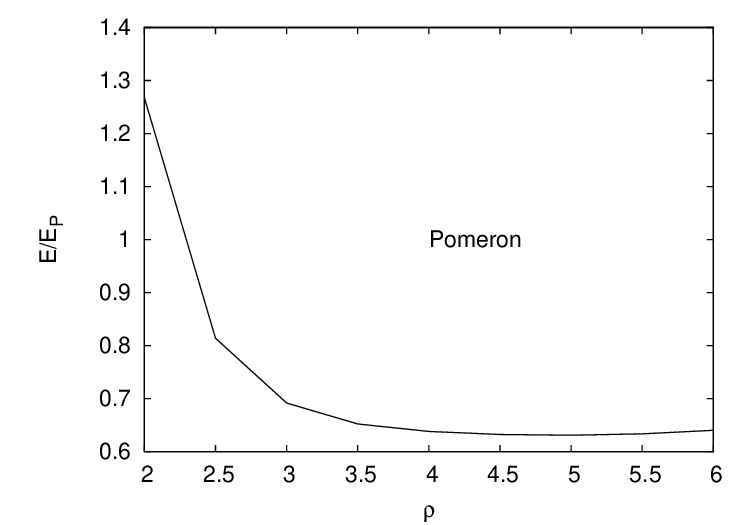, width=8 cm}
\caption{The pomeron ground state energies $E$ vs their theoretical values without odderon $E_P$
for different $\rho$ in the model with odderon.}
\end{center}
\label{fig2}
\end{figure}
We observe that the ground state energy  becomes lower due to interaction with odderons, being 2/3 of $E_P$ at high $\rho$ (low $\lambda$).
Note that at $\rho<3$  Eq.~(\ref{ep}) poorly describes the actual ground state energy level  in absence of odderons, the latter being considerably greater.
This has to be taken into account comparing the levels with and without odderons from Fig. 2.

\section{Two transverse dimensions}
Passing to the real world we are to consider the model with $D=2$. However, the renormalization technique that we are going to apply
requires a model with $D=4$. So as usual in such cases we start with the theory in $D=4-\epsilon$ dimensions, study it in the vicinity of small positive $\epsilon$ and
afterwards continue to $\epsilon=2$.

\subsection{Renormalization}
The initial Lagrangian  is assumed to be written in unrenormalized fields and parameters. So its form is (\ref{eq1}) with
$\phi_i\to\phi_{i0}$, $\bphi_i\to\bphi_{i0}$ $\alpha'_i\to\alpha'_{i0}$, $\lambda_l\to\lambda_{l0}$, where $i=1,2$ and $l=1,2,3$.
Let the lowest
 eigenvalue  of the Hamiltonian for the pomeron and odderon sectors be $M_i(\mu_{10},\mu_{20})$, $i=1,2$.
At these energies  the inverse propagators $\Gamma_i(E,k^2)$ vanish
\beq
\Gamma_i(E,k^2)|_{E=M_i(\mu_{10},\mu_{20}),k=0}=0,\ \ i=1,2.
\label{eq4}
\eeq
We assume that  $M_1(\mu_{10},\mu_{20})$, initially positive, diminishes as $\mu_{10}$
grows up to its maximal value
$\mu_{10c}$ at which $M_1$ reaches its critical value $M_{1c}=0$ compatible with the Froissart bound, as occurs in the perturbative approach.
This suggests introducing instead of  $\mu_{10}$  variable
$\delta_{10}=\mu_{10c}-\mu_{10}$,
which is initially non-negative and vanishes when $\mu_{10}$ and $M_1$ attain their critical values at fixed $\mu_{20}$.
Note that in the free theory with $\lambda_{i0}=0$ we  have $M_1=\mu_{10}$.
It becomes equal to zero at $\mu_{10}=0$. As a result, in the free theory
$\mu_{10c}=0$, which means that in the presence of interaction $\mu_{10c}$ is of the second order
in $\lambda_0$ and corresponds to mass renormalization.
Similarly for $M_2(\mu_{10},\mu_{20})$ we determine the value $\mu_{20c}$ at which $M_2$ attains its minimal value $M_{2c}=0$
at fixed $\mu_{10}$ and define a non-negative variable $\delta_{20}$ as the difference
$\delta_{20}=\mu_{20c}-\mu_{20}$.

The unrenormalized inverse full propagators acquire the form
\beq
\Gamma_j(E,k^2)=E-\delta_{j0}-\alpha'_{j0}k^2+\mu_{j0c}-\Sigma_j(E,k^2),\ \ j=1,2,
\label{m0}
\eeq
where $\Sigma_j$ are the non-trivial self-masses. In the lowest approximation they are
graphically shown in Fig. 3.
\begin{figure}
\begin{center}
\epsfig{file=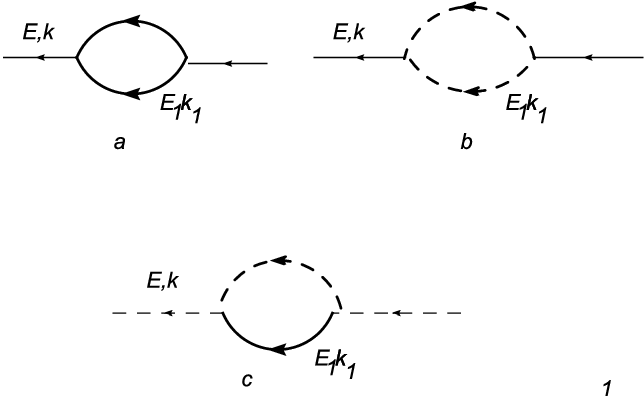, width=8 cm}
\caption{Self masses for $\Gamma_1$ ($a+b$) and $\Gamma_2$ ($c$). Pomerons and odderons are shown by solid and dashed lines, respectively.}
\end{center}
\label{fig3}
\end{figure}
From
 (\ref{eq4}) we find
\beq
\mu_{j0c}=\Sigma_j(E,k^2)_{E=k=\delta_j=0},\ \ j=1,2.
\label{mu1c}
\eeq
So  denoting
\[\Sigma_j(E,k^2)-\Sigma_j(E,k^2)_{E=k=\delta_j=0}=\tilde{\Sigma}_j(E,k^2),\ \ j=1,2\]
we obtain the unrenormalized inverse propagators as
\beq
\Gamma_j=E-\alpha'_{j0}k^2-\delta_{j0}-\tilde{\Sigma}_j.
\label{gamunren}
\eeq

Renormalized quantities are introduced in the standard manner:
\[\phi_i=Z_i^{-1/2}\phi_{i0},\ \
\alpha'_i=U_{i}^{-1}Z_i\alpha'_{i0},\ \
\delta_i=T_i^{-1}Z_i\delta_{i0},\ \ i=1,2,\]
\[\lambda_1=W_{1}^{-1}Z_{1}^{3/2}\lambda_{10},\ \ \lambda_{2,3}=W_{2,3}^{-1}Z_1^{1/2}Z_2\lambda_{20,30}.\]

The generalized vertices for transition  of $n_1$ pomerons and $n_2$  odderons to $m_1$ pomerons and $m_2$ odderons
transform as
\[\Gamma^{R,n_1,n_2,m_1,m_2}(E_i,k_i,E_N)=Z_1^{(n_1+m_1)/2}Z_2^{(n_2+m_2)/2}
\Gamma^{n_1,n_2,m_1,m_2}(E_i,k_i),\]
where $E_N$ is the renormalization energy point.

Constants $Z$, $U$ $T$ and $W$ are determined by the renormalization conditions imposed on renormalized quantities
\[\frac{\pd}{\pd E}\Gamma_i^R(E, k^2, E_N)\Big|_{E=-E_N,k^2=\delta_j=0}=1,\ \ i,j=1,2,\]\[
\frac{\pd}{\pd k^2}\Gamma_i^R(E, k^2, E_N)\Big|_{E=-E_N,k^2=\delta_j=0}=-\alpha'_i,\ \ i,j=1,2,\]\beq
\frac{\pd}{\pd \delta_i}\Gamma_i^R(E, k^2, E_N)\Big|_{E=-E_N,k^2=\delta_j=0}=-1,\ \ i,j=1,2,
\label{renormc}
\eeq
\[\Gamma^{R,1,0,2,0}(E_i,k_i,E_N)\Big|_{E_1=2E_2=2E_3=-E_N, k_j=\delta_j=0}=i\lambda_1 (2\pi)^{-(D+1)/2},
\ \ j=1,2,\]
\[\Gamma^{R,0,1,1,1}(E_i,k_i,E_N)\Big|_{E_1=2E_2=2E_3=-E_N, k_j=\delta_j=0}=i\lambda_2 (2\pi)^{-(D+1)/2},
\ \ j=1,2,\]
\[\Gamma^{R,1,0,0,2}(E_i,k_i,E_N)\Big|_{E_1=2E_2=2E_3=-E_N, k_j=\delta_j=0}=i\lambda_3 (2\pi)^{-(D+1)/2},
\ \ j=1,2.\]

We introduce new dimensionless coupling constants: unrenormalized  $u_0\!$ and renormalized $u\!$
\[ g_{40}\equiv u_0=\frac{\alpha'_{20}}{\alpha'_{10}},\ \ g_4\equiv u=\frac{\alpha'_{2}}{\alpha'_1}.\]
With these normalizations the renormalization constants $Z$, $U$ $T$ and $W$  depend only on the dimensionless coupling constants
\beq
 g_i=\frac{\lambda_i}{(8\pi\alpha'_1)^{D/4}E_N^{(4-D)/4}},\ \ i=1,2,3\ \ {\rm and}\ \  g_4\equiv u.
 \label{gi}
\eeq

The RG equations  are standardly obtained from the condition that the unrenormalized $\Gamma$ do not depend on $E_N$.
So differentiating $\Gamma^R$ with respect to $E_N$ at $\lambda_{i0}, u_0,\alpha'_{10},\delta_{j0}$  fixed we get
\beq
\Big(E_N\frac{\pd}{\pd E_N}+\sum_{i=1}^4\beta_i(g)\frac{\pd}{\pd g_i}+
\sum_{i=1}^{2}\kappa_i(g)\delta_i\frac{\pd}{\pd \delta_i}+\tau_1(g)\alpha'_1\frac{\pd}{\pd\alpha'_1}
-\sum_{i=1}^2\frac{1}{2}(n_i+m_i)\gamma_i(g)\Big)\Gamma^{R}=0,
\label{eq44}
\eeq
where
\[
\beta_i(g)=E_N\frac{\pd g_i}{\pd E_N},\ \ i=1,...,4,\]
\[\gamma_i(g)=E_N\frac{\pd \ln Z_i}{\pd E_N},\ \
\tau_i(g)=E_N\frac{\pd}{\pd E_N}\ln\Big(U_{i}^{-1}Z_i\Big),\ \
\kappa_i(g)=E_N\frac{\pd}{\pd E_N}\ln\Big(T_{i}^{-1}Z_1\Big),\ \ i=1,2\]
and the derivatives are taken at $\lambda_{i0}$, $u_0$, $\alpha'_{10}$ and $\delta_{i0}$ fixed.
For brevity  we denote in the following
\beq
\gamma(g)=\sum_{i=1}^2\frac{1}{2}(n_i+m_i)\gamma_i(g).\label{gab}\eeq

\subsection{Anomalous dimensions, beta-functions and fixed points}
The renormalized inverse propagators are
\beq
\Gamma_j^R=Z_j\Gamma_j
=Z_jE-U_j\alpha'_{j}k^2-T_j\delta_j-\tilde{\Sigma}_j(E,k^2),
\label{gamren}
\eeq
where we put $Z_j=1$ in front of $\Sigma_j$ having in mind the lowest non-trivial order.
We can rewrite (\ref{gamren}) as
\beq
\Gamma_j^R=E-\alpha'_jk^2-\delta_j-\Sigma^R_j(E,k^2),\ \ j=1,2,\label{gamr}\eeq
where
\beq
\Sigma_j^R=\tilde{\Sigma}_j-(Z_j-1)E+(U_j-1)\alpha'_j k^2+(T_j-1)\delta_j.
\label{sigr}
\eeq

Calculation of diagrams in Fig. 3 leads to the following expressions for $\Sigma$.
For the pomeron $\Sigma_1=\Sigma_1^a+\Sigma_1^b$, where
\beq
\Sigma_1^a=\frac{1}{2}g_1^2E_N^{2-D/2}\Gamma(1-D/2)\Big(\frac{1}{2}\alpha'_{1}k^2-E+2\delta_1\Big)^{D/2-1},
\label{sig1a}
\eeq
\beq
\Sigma_1^b=-\frac{1}{2}\frac{g_3^2E_N^{2-D/2}}{u^{D/2}}\Gamma(1-D/2)\Big(\frac{1}{2}\alpha'_2k^2-E+2\delta_2\Big)^{D/2-1}
\label{sig1b}
\eeq
and
\beq
\Sigma_2=\frac{g_2^2E_N^{2-D/2}}{[(1+u)/2]^{D/2}}\Gamma(1-D/2)\Big(\frac{\alpha'_2 k^2}{1+u}-E+\delta_1+\delta_2\Big)^{D/2-1}
\label{sig2}
\eeq
for the odderon.
To build the renormalized $\Sigma_i^R$ one needs $Z_i,U_i$ and $T_i$. They are determined by the derivatives
\[Z_i-1=\frac{\pd}{\pd E}\tilde{\Sigma}_i(E,k^2)_{E=-E_N,k=\delta_1=\delta_2=0},\]
\[(U_i-1)\alpha'_i=-\frac{\pd}{\pd k^2}\tilde{\Sigma}_i(E,k^2)_{E=-E_N,k=\delta_1=\delta_2=0},\].
\[T_i-1=-\frac{\pd}{\pd \delta_i}\tilde{\Sigma}_i(E,k^2)_{E=-E_N,k=\delta_1=\delta_2=0}.\]
Calculations give for the pomeron
\[Z_1^a-1=\frac{1}{2}g_1^2\Gamma(2-D/2),\ \ U_1^a-1=\frac{1}{4}g_1^2\Gamma(2-D/2),\ \ T_1^a-1=g_1^2\Gamma(2-D/2),\]
\[Z_1^b-1=-\frac{1}{2}\frac{g_3^2}{u^{D/2}}\Gamma(2-D/2),\ \ U_1^b-1=-\frac{1}{4}\frac{g_3^2}{u^{D/2}}u\Gamma(2-D/2),\ \ T_1^b-1=0\]
and for the odderon
\[Z_2-1=\frac{g_2^2}{[(1+u)/2]^{D/2}}\Gamma(2-D/2),\ \ U_2-1=\frac{g_2^2}{[(1+u)/2]^{D/2}}\frac{1}{1+u}\Gamma(2-D/2),\]
\[T_2-1=\frac{g_2^2}{[(1+u)/2]^{D/2}}\Gamma(2-D/2).\]
One can check that with these renormalization constants the renormalized masses $\Sigma_i^R$ turn out to be finite.

To find the anomalous dimensions we have to differentiate the renormalization constants over $E_N$.
In the lowest order we have for all constants
\[\frac{\pd}{\pd E_N}\ln Z=\frac{\pd}{\pd E_N}\ln (1+Z-1)=\frac{\pd}{\pd E_N}(Z-1).\]
All renormalization constants depend on $E_N$
via constants $g_i$, $i=1,2,3,4$, which in the lowest order are equal to the unrenormalized $g_{i0}$.
Since  $E_N\pd g_i^2/\pd E_N=(D/2-2)g_i^2$, $i=1,2,3$
and $g_4=u$ does not depend on $E_N$ in this order,
to find the anomalous dimensions we have only to multiply the renormalization constants by $(D/2-2)$.
We find in the lowest order in small $\epsilon$
\beq
\gamma_1^a=-\frac{1}{2}g_1^2,\ \
\gamma_1^b=\frac{1}{2}\frac{g_3^2}{u^2},\ \
\gamma_2=-\frac{4g_2^2}{(1+u)^2},\label{gamma}\eeq
\beq
\tau_1^a=-\frac{1}{4}g_1^2,\ \ \tau_1^b=\frac{1}{4}\frac{g_3^2}{ u^2}(2-u),\ \
\tau_2=-\frac{4ug_2^2}{(1+u)^3},\label{tau}\eeq
\beq
\kappa_1^a=\frac{1}{2}g_1^2,\ \
\kappa_1^b=\frac{1}{2}\frac{g_3^2}{u^2},\ \
\kappa_2=0.
\label{kapp2}
\eeq

To calculate $\beta$-functions one has to calculate the relevant diagrams for the non-trivial couplings.
In the single loop approximation which is in our scope we have to calculate the relevant triangle diagrams
shown in Fig. 4.
\begin{figure}[h!p]
\begin{center}
\includegraphics[width=12 cm]{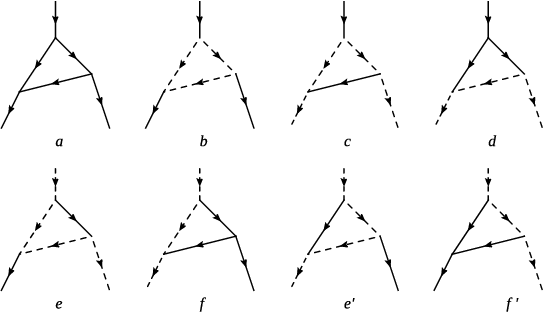}
\caption{Diagrams for $\Gamma^{10,20}$ ($a+b$), $\Gamma^{10,02}$ ($c+d$) and $\Gamma^{01,11}$ ($e+f$).
Inverse diagrams for $\Gamma^{10,20}$ and  $\Gamma^{10,02}$ are identical.
For $\Gamma^{01,11}$ inverse diagrams are different and shown as ($e'$) and ($f'$).
Pomerons and odderons are shown by solid and dashed lines, respectively.}
\label{fig4}
\end{center}
\end{figure}

In the lowest order in  small $\epsilon$ the $\beta$-functions are found in ~\cite{bkv2023} to be the following.
If $u\neq 0$ the four $\beta$-functions are
\beq
\beta_1=-\frac{1}{4}\epsilon g_1+\frac{3}{2}g_1^3- g_2g_3^2\frac{2}{u^2}+g_1g_3^2\frac{1+u}{4u^2},
\label{bet1}
\eeq
\beq
\beta_2
=-\frac{1}{4}\epsilon g_2+g_1g_2^2\frac{6+2u}{(1+u)^2}- g_2g_3^2\frac{1+8u-u^2}{4u^2(1+u)},
\label{bet2}
\eeq
\beq
\beta_3
=-\frac{1}{4}\epsilon g_3+g_1g_2g_3\frac{4}{1+u}+g_2^2g_3\frac{4}{u(1+u)^2}+ g_3^3\frac{u-1}{4u^2},
\label{bet3}
\eeq
and
\beq
\beta_4=g_1^2\frac{u}{4}-g_2^2\frac{4u^2}{(1+u)^3}+ g_3^2\frac{u-2}{4u}.
\label{bet4}
\eeq

If $u=g_4=0$ one has to pass  from $g_3$ to a new coupling constant $r=g_3/g_4$
and the 4-dimensional domain of coupling constants $g_i$, $i=1,...,4$ splits into two 3-dimensional
domains with either $r=0$ or $g_2=0$.

If the initial $g_2=0$ then $\bar{g}_2(t)=0$ and $g_1,\ r$ and $g_4$ evolve with $\beta$-functions
\beq
\beta_1=-\frac{1}{4}\epsilon g_1+\frac{3}{2}g_1^3+g_1r^2\frac{1+u}{4},
\label{bet1b}
\eeq
\beq
\beta_r=r\Big(-\frac{1}{4}\ep-g_1^2\frac{1}{4}+ \frac{1}{4}r^2\Big),\ \
\beta_4=g_1^2\frac{u}{4}+ r^2\frac{u(u-2)}{4}.
\label{bet4b}
\eeq

If the intial $r=0$ then $\bar{r}(t)=0$ and $g_1,\ g_2$ and $g_4$ evolve with $\beta$-functions
\beq
\beta_1=-\frac{1}{4}\epsilon g_1+\frac{3}{2}g_1^3,\ \
\beta_2
=-\frac{1}{4}\epsilon g_2+g_1g_2^2\frac{6+2u}{(1+u)^2},
\label{bet2c}
\eeq
\beq
\beta_4=g_1^2\frac{u}{4}-g_2^2\frac{4u^2}{(1+u)^3}.
\label{bet4c}
\eeq

The fixed points are determined by conditions $\beta_i=0$, $i=1,...,4$.
We have found five real fixed points (FP)
presented in the left Table in which we show the
corresponding coupling constants at the fixed point
with $g_{1,2,3}$ divided by $\sqrt{\ep}$.

\begin{center}
{\bf Table }\\
\vspace{0.2cm}
Left: Coupling constants at fixed points with $g_{1,2,3}$ divided by $\sqrt{\ep}$.\\
Right: Eigenvalues of  matrix $2a$ at $\ep=2$.\\
\vspace{0.2cm}
\begin{tabular}{|r|r|r|r|r|}
\hline
FP&$g_1$&$g_2$&$g_3$&$g_4$\\\hline
$g_g^{(0)}$&0&0&$0$&0\\\hline
$g_c^{(1)}$&$\frac{1}{\sqrt{6}}$&0&0&0\\\hline
$g_c^{(2)}$&$\frac{1}{\sqrt{6}}$&0.40&0&0.89\\\hline
$g_c^{(3)}$&$\frac{1}{\sqrt{6}}$&$\frac{1}{\sqrt{96}}$&0&0\\\hline
$g_c^{(4)}$&0&0&2&2\\
\hline
\end{tabular}
\hspace {1 cm}
\begin{tabular}{|r|r|r|r|r|}
\hline
FP&$x_1$&$x_2$&$x_3$&$x_4$\\\hline
$g_c^{(0)}$&0&-2&-2&0\\\hline
$g_c^{(1)}$&2&-1&-1&1/6\\\hline
$g_c^{(2)}$&2&1.21&0.37&-0.14\\\hline
$g_c^{(3)}$&2&1&1/6&1/(6u)\\\hline
$g_c^{(4)}$&2&-16/3&2&2\\
\hline
\end{tabular}
\end{center}
At $g_c^{(0)}$ ratio $r=g_3/g_4=1$, at $g^{(1)}$ and at $g_c^{(3)}$ $r=0$.

Attraction or repulsion at the fixed points is described by the matrix $a_{ij}=\partial \beta_i/\partial g_j$, $i,j=1,...,4$.
Its eigenvectors for positive (negative) eigenvalues indicate direction along which the fixed point is attractive (repulsive).
In the right Table we show  eigenvalues $x=\{x_1,x_2, x_3,x_4\}$ for  matrix $2a$ at $\ep=2$.
Zero eigenvalues describe directions along which the corresponding projection of the 4-vector $g_i$ does not move
in the vicinity of the fixed point and stays equal to its initial value.
Of all fixed points only $g_c^{(3)}$ is purely attractive. All the rest have one or several repulsive
directions, so that arriving at them is only possible from a restricted domain of all coupling constants.
This is confirmed by numerical tests.

\section{Evolution}
From the dimensional analysis we get
\[[\phi_i]=[\bphi_i]=k^{D/2},\ \ [\alpha'_i]=Ek^{-2},\ \ [\delta_i]=E,\ \ i=1,2,\]
\[
\Big[\Gamma^{R}\Big]=Ek^{D-(n+m)D/2},\ \ n=n_1+n_2,\ \ m=m_1+m_2.\]
This allows to write
\beq
\Gamma^{R}(E_i,k_i,g,\alpha'_1,\delta_{1,2},E_N)=E_N\Big(\frac{E_N}{\alpha'_1}\Big)^{(2-n-m)D/4}
\Phi\Big(\frac{E_i}{E_N},\frac{\alpha'_1}{E_N}{k_i}{k_j}, \frac{\delta_{1,2}}{E_N},g\Big).
\label{scfun}
\eeq.

Using the scale transformation
$E\to E/\xi,\ \ k\to k$
we find from the scale invariance
\beq
\Gamma^{R}(E_i,k_i,g,\alpha'_1,\delta_{1,2},E_N)=
\xi\Gamma^{R}\Big(\frac{E_i}{\xi},k_i,g,\frac{\alpha'_1}{\xi},\frac{\delta_{1,2}}{\xi},\frac{E_N}{\xi}\Big).
\label{gamxi}
\eeq
Combining this equation with  the RG equation (\ref{eq44}) one gets evolution equations for energy $E$ or masses $\delta_{1,2}$.

\subsection{Evolution of energy $E$}
Putting $E\to\xi E$ in (\ref{gamxi}) we find
\beq
\Gamma^{R}(\xi E_i,k_i,g,\alpha'_1,\delta_{1,2},E_N)=
\xi\Gamma^{R}\Big(E_i,k_i,g,\frac{\alpha'_1}{\xi},\frac{\delta_{1,2}}{\xi},\frac{E_N}{\xi}\Big).
\label{gamxia}
\eeq
Differentiating (\ref{gamxia})  in $\xi$ and using the RG equation we arrive at
\[
\Big\{\xi\frac{\pd}{\pd\xi}-\sum_{i=1}^4\beta_i(g)\frac{\pd}{\pd g_i}+[1-\tau_1(g)]\alpha'_1\frac{\pd}{\pd\alpha'_1}
+\sum_{i=1,2}[1-\kappa_i(g)]\delta_i \frac{\pd}{\pd \delta_i}\]\beq
-[1-\gamma(g)] \Big\}\Gamma^R(\xi E_i,k_i,g,\alpha'_1,\delta_{1,2},E_N)=0.
\label{old055}
\eeq
The solution of this equation is standard
\[
\Gamma^R(\xi E_i,{k}_i,g,\alpha'_1,E_N)=\Gamma^R\Big(E_i,{k}_i,
\bg(-t),\balpha'_1(-t),\bar{\delta}_{1,2}(-t),E_N\Big)\]
\beq
\times
\exp\Big\{\int_{-t}^0dt' \Big[1-\gamma(\bg(t'))\Big]\Big\},
\label{eq056}
\eeq
where
\beq
\frac{d \bg_i(t)}{dt}=-\beta_i(\bg(t)),\ \
\frac{d\ln\balpha'_1(t)}{dt}=1-\tau_1(\bg(t)),\ \
\frac{d\ln\bar{\delta}_i}{dt}=1-\kappa_i(\bg(t))
\label{ddelta}\eeq
and we use (\ref{gab}).
The initial conditions are
\beq
\bg_i(0)=g_i,\ \ \balpha'_1(0)=\alpha'_1,\ \ \bar{\delta}_{1,2}(0)=\delta_{1,2}
\label{inic}
\eeq
and we denote
$t=\ln\xi$.

At the fixed point $g_i=g_{ic}$ we have
  $\bg_i(t)=g_{ic}$  fixed and
\beq
\balpha'_1(-t)=\alpha'_1 e^{-tz},\ \ z=1-\tau_1(g_c),\ \
\bar{\delta}_{1,2}=\delta_{1,2}e^{-t\zeta_{1,2}},\ \
\zeta_{1,2}=1-\kappa_{1,2}(g_c).
\label{balpha}\eeq
The solution (\ref{eq056}) at $g=g_c$ becomes
\beq
\Gamma^R(\xi E_i, k_i,\alpha'_1,\delta_{1,2},E_N)=
\Gamma^R(E_i,k_i,\alpha'_1 e^{-zt},\delta_{1,2}E^{-\zeta_{1,2}t},E_N)e^{t[1-\gamma(g_c)]} .
\label{eq084}
\eeq

We use the scaling property
\[\Gamma^R(E_i,k_i,\alpha'_1,\delta_{1,2},E_N)=E_N\Big(\frac{E_N}{\alpha'_1}\Big)^{(2-n-m)D/4}\Phi\Big(\frac{E_i}{E_N},
\frac{\alpha'_1}{E_N}{k}_i{k}_j,\frac{\delta_{1,2}}{E_N}\Big)\]
together with (\ref{eq084})
to obtain at $E\to E/\xi$
\[\Gamma^R(E_i,k_i,\alpha'_1,\delta_{1,2},E_N)=\]\[e^{t[1-\gamma(g_c)]}E_N\Big(\frac{E_N}{\alpha'_1}\Big)^{(2-n-m)D/4}e^{tz(2-n-m)D/4}
\Phi\Big(\frac{E_i}{E_N\xi},\frac{\alpha'_1 e^{-zt}}{E_N}{k}_i{k}_j,\delta_{1,2}e^{-\zeta_{1,2}t}\Big) .\]
Taking
$\xi=-E/E_N$
where $E$ is the sum of energies in the final state $m=\{m_1,m_2\}$,
we find finally
\[\Gamma^R(E_i,k_i,\alpha'_1,\delta_{1,2},E_N)=E_N\Big(\frac{E_N}{\alpha'_1}\Big)^{(2-n-m)D/4}
\Big(\frac{-E}{E_N}\Big)^{1-\gamma(g_c)+z(2-n-m)D/4}\]\beq\times
\Phi\Big(\frac{-E_i}{E},\alpha'_1\frac{{k}_i{k}_j}{E_N}e^{-zt},
\delta_{1,2}e^{-\zeta_{1,2}t}\Big).\label{gamrphi}\eeq
In particular, we find
\beq
\Gamma_i(E,k^2,g_c,\alpha'_1,E_N)=E_N\Big(\frac{-E}{E_N}\Big)^{1-\gamma_i(g_c)}\Phi_i(\rho,\rho_1,\rho_2),
\ \ i=1,2,
\label{eq087}
\eeq
where
\beq\rho=\Big(\frac{-E}{E_N}\Big)^{-z}\,\frac{\alpha'_1k^2}{E_N},\ \
\rho_{1,2}=\delta_{1,2}\Big(\frac{-E}{E_N}\Big)^{-\zeta}.
\eeq
Inverse propagators $\Gamma_i$ have a zero when
$\Phi_i(\rho,\rho_1,\rho_2)=0$, which determines the Regge trajectory.
In particular, in the limiting case $\delta_1=\delta_2=0$ we find
 the trajectory as
\beq
\alpha_i(k^2)=1+E_N\Big(\frac{\alpha'_1 k^2}{E_N}\Big)^{1/z}f_i(g_c),\ \ i=1,2.
\label{eq089}
\eeq
Generally, it  is not analytic at $k^2=0$.  In the model without odderon the slope is infinite at $k^2=0$  ~\cite{abarb1}.

The actual behaviour depends on the choice of the fixed point.
For the most important fully attractive fixed point $g_c^{(3)}$ we find at $\epsilon=2$
\[ \gamma_1=-\frac{1}{6},\ \ \tau_1=-\frac{1}{12},\ \ \kappa_1=\frac{1}{6},\ \
\gamma_2=-\frac{1}{12},\ \ \tau_2=0,\ \ \kappa_2=0.\]
So we get the inverse propagators
\beq \Gamma^R_1=-E\Big(\frac{-E}{E_N}\Big)^{1/6}\Phi_1(\rho,\rho_1,\rho_2),\ \
\Gamma^R_2=-E\Big(\frac{-E}{E_N}\Big)^{1/12}\Phi_2(\rho,\rho_1,\rho_2)\label{gam3},\eeq
where
\[\rho=\Big(\frac{-E}{E_N}\Big)^{-13/12}\frac{\alpha'_1 k^2}{E_N},\ \
\rho_1=\Big(\frac{-E}{E_N}\Big)^{-5/6}\delta_1,\ \
\rho_2=\Big(\frac{-E}{E_N}\Big)^{-1}\delta_2.\]
The propagators $G$ as functions of energy squared $s$ are obtained by the inverse Mellin transform
and have the asymptotical behaviour at $s\to\infty$
\beq
G_1(s)\sim s(\ln s)^{1/6},\ \ G_2(s)\sim s(\ln s)^{1/12}.
\label{prop}\eeq
We postpone the discussion of the elastic scattering amplitude to the next Section, since it  involves coupling to participants.

Note that comparing (\ref{eq087}) to the explicit expressions for the inverse propagators in the first two orders
in small $\epsilon$ one can find an explicit expression for the scaling function $\Phi$. For the model without odderons this procedure was
realized in ~\cite{abarb1,abarb2}.

\subsection{Evolution of masses $\delta_{1,2}$}
To study evolution of $\delta_1$
we put  $\delta_1\to\xi \delta_1$ in (\ref{gamxi}).
Differentiating the result   in $\xi$ and using the RG equation we arrive at
\[
\Big([1-\kappa_1(g)]\xi\frac{\pd}{\pd\xi}-\sum_{i=1}^4\beta_i(g)\frac{\pd}{\pd g_i}+[1-\tau_1(g)]\alpha'_1\frac{\pd}{\pd\alpha'_1}\]
\[+[1-\kappa_2(g)]\delta_2\frac{\pd}{\pd\delta_2}
-E\frac{\pd}{\pd E}-[1-\gamma(g)]\Big)\Gamma^R(E_i,k_i,g,\alpha'_1,\xi\delta_1,\delta_2,E_N)=0.\]
We put
$t=\ln\xi,\ \ d(\bar{g}(t))=1-\kappa_1(\bar{g}(t))$.
Then the solution is
\[
\Gamma^R(E_i,k_i,g,\alpha'_1,\xi\delta_1,\delta_2,,E_N)=
\Gamma^R\Big(\bar{E}_i(-t),k_i,\bar{g}(-t),\bar{\alpha}'_1(-t),\bar{\delta}_2(-t),\delta_1,E_N\Big)\]
\beq
\times
\exp\Big\{\int_{-t}^0dt'\frac{1-\gamma(\bar{g}(t'))}{1-\kappa_1(\bar{g}(t'))}\Big\},
\label{eq54}
\eeq
where
\beq
\frac{d \bar{g}_i(t)}{dt}=-\frac{\beta_i(\bar{g}(t))}{d(t)},\ \
\frac{d\ln\bar{\alpha}'_1(t)}{dt}=\frac{1-\tau_1(\bar{g}(t))}{d(t)},\ \
\frac{d\ln\bar{\delta}_2(t)}{dt}=\frac{1-\kappa_2(\bar{g}(t))}{d(t)},\ \
\frac{d\ln\bar{E}_i(t)}{dt}=\frac{1}{d(t)}
\label{eq55}
\eeq
with the initial conditions
\[\bar{g}_i(0)=g_i,\ \ \bar{\alpha}_1'(0)=\alpha'_1,\ \ \bar{\delta}_2(0)=\delta_2,\ \ \bar{E}_i(0)=E_i.\]

At the fixed point $\bg_i(t)=g_{ic}$ (using that $\kappa_2=0$ in this order)
\beq
\bar{E}_i(-t)=E_i e^{-t\zeta},\ \
\balpha'_1(-t)=\alpha'_1e^{-tz},\ \
\bdel_2(-t)=\delta_2e^{-t\zeta},\eeq
where
\[\zeta=\frac{1}{d(g_c)},\ \ z=\frac{1-\tau_1(g_c)}{d(g_c)}.\]
Combining
the solution (\ref{eq54}) at $g=g_c$  and scaling property, similarly to the
previous subsection, we arrive at
\[\Gamma^R(E_i,k_i,g_c,\alpha'_1,\delta_1,\delta_2,E_N)=\]\[CE_N\Big(\frac{E_N}{\alpha'_1}\Big)^{(2-n-m)D/4}
\Phi\Big(\frac{E_i}{E_N,}e^{-t\zeta},\frac{\alpha'_1}{E_N}{k_i}{k_j}e^{-tz},\frac{\delta_2}{E_N}e^{-t\zeta}\Big),\]
where
\[C=\exp\Big(t[1-\gamma(g_c)]/d(g_c)+tz(2-n-m)D/4\Big).\]
In particular,
taking $\xi=\delta_1/E_N$,
we find for the inverse full propagators
\beq
\Gamma^R_j(E,k^2,g_c,\alpha'_1,\delta_1,\delta_2,E_N)=\delta_1\Big(\frac{\delta_1}{E_N}\Big)^{-\tilde{\gamma}_j}      
\Phi_j(\rho_1,\rho_2,\rho_3),
\ \ j=1,2,
\label{eq087a}
\eeq
where $\tilde{\gamma}_j=(\gamma_j-\kappa_1)/d$.
Here $\rho_i$ are given by
\beq
\rho_1=\frac{E}{E_N}e^{-t\zeta},\ \ \rho_2=\frac{\alpha'_1}{E_N}k^2e^{-tz},\ \ \rho_3=\frac{\delta_2}{E_N}e^{-t\zeta}.
\label{rho}
\eeq

The actual behaviour again depends on the choice of the fixed point.
For the most important fully attractive fixed point $g_c^{(3)}$ we find at $\epsilon=2$
\[\tilde{\gamma_1}=-\frac{2}{5},\ \ \tilde{\gamma}_2=-\frac{3}{10},\ \ \zeta-1=\frac{1}{5},\ \ z-1=\frac{3}{10}.\]
So at small $\delta_1$ the inverse propagators are
\beq
\Gamma^R_1=\delta_1\Big(\frac{\delta_1}{E_N}\Big)^{2/5}\Phi_1(\rho_1,\rho_2,\rho_3),\ \
\Gamma^R_2=\delta_1\Big(\frac{\delta_1}{E_N}\Big)^{3/10}\Phi_2(\rho_1,\rho_2,\rho_3),\label{gam3a}\eeq
where
\[\rho_1=\frac{E}{\delta_1}\Big(\frac{\delta_1}{E_N}\Big)^{-1/5},\ \
\rho_2=\frac{\alpha'_1k^2}{\delta_1}\Big(\frac{\delta_1}{E_N}\Big)^{-3/10},\ \
\rho_3=\frac{\delta_2}{\delta_1}\Big(\frac{\delta_1}{E_N}\Big)^{-1/5}.\]

The study of small $\delta_2$ is done in completely the same manner interchanging in the derivation $\delta_1\leftrightarrow \delta_2$.
One has to take into account that $\kappa_2=0$ and so denominator $d=1$.
At the fixed point $g_c^{(3)}$ we have
\[\tilde{\gamma_1}=-\frac{1}{6},\ \ \tilde{\gamma}_2=-\frac{1}{12},\ \ \zeta-1=-\frac{1}{6},\ \ z-1=\frac{1}{12}.\]
So at small $\delta_2$ the inverse propagators are
\beq
\Gamma^R_1=\delta_2\Big(\frac{\delta_2}{E_N}\Big)^{1/6}\Phi_1(\rho_1,\rho_2,\rho_3),\ \
\Gamma^R_2=\delta_2\Big(\frac{\delta_2}{E_N}\Big)^{1/12}\Phi_2(\rho_1,\rho_2,\rho_3)\label{gam3b}\eeq
where
\[\rho_1=\frac{E}{\delta_2},\ \ \rho_2=\frac{\alpha'_1k^2}{\delta_2}\Big(\frac{\delta_2}{E_N}\Big)^{-1/12},\ \
\rho_3=\frac{\delta_1}{\delta_2}\Big(\frac{\delta_2}{E_N}\Big)^{1/6}.\]


\section{Elastic scattering amplitude}
\subsection{The asymptotic }
We consider the elastic scattering of two particles with exchanges of  pomerons and odderons.
It is the sum of contributions ${\cal A}^{(nm)}(s,t)$ in which the projectile emits $n_1$ pomerons and $n_2$ odderons and the target absorbs
$m_1$ pomerons and $m_2$ odderons.
Here $(nm)=(n_1,n_2,m_1,m_2)$ where $n_1$ and $n_2$ are numbers of incoming pomerons and odderons and
$m_1$ and $m_2$ are numbers of outgoing pomerons and odderons, respectively. In the following we denote
the number of initial reggeons (pomeron plus odderons) $n=n_1+n_2$, the number of final reggeons $m=m_1+m_2$, the total number of pomerons
(initial plus final) $n_P=n_1+m_1$, the total number of odderons $n_O=n_2+m_2$. The total number of all reggeons is evidently
$n_t=n+m=n_P+n_O$.
Amplitude ${\cal A}$ with a given number of exchanged reggeons is shown in Fig. 5.
\begin{figure}
\begin{center}
\epsfig{file=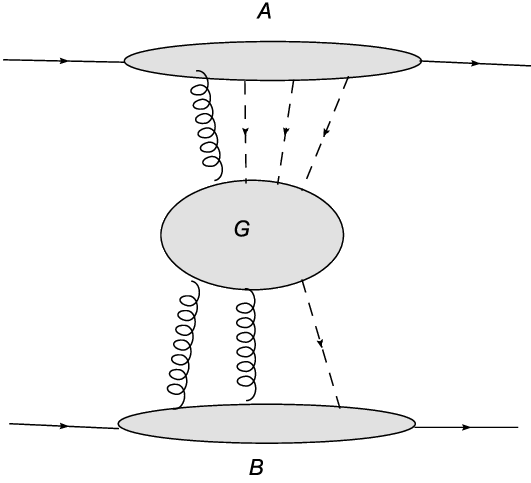, width=8 cm}
\caption{Elasic amplitude with a given number of pomerons (solid lines) and odderons (dashed lines) exchanges.}
\end{center}
\label{fig5}
\end{figure}
For simplicity we assume that the couplings of the reggeons  to the participants are just (unknown) constants,
namely $A^{n_1,n_2}$ and $B^{m_1,m_2}$.
This roughly speaking corresponds to the Glauber coupling. In this case the Mellin-transformed amplitude, that is
in the complex angular momentum space variables, will be given by the integral over all internal energetic and momentum
integration variables
\[
{\cal A}^{(n,m)}(E,q)=A^{n_1,n_2}B^{m_1,m_2} I^{(n,m)}(E,t),\]
where
\[
I^{(nm)}(E,t)=\]\beq=\int\prod_{i=1}^{n+m}d^Dk_idE_i
\delta^D(\sum_{in}k_i-q)\delta^D(\sum_{out}k_i-q)
\delta(\sum_{in}E_i-E)\delta(\sum_{out}E_i-E)
G_R^{(nm)}(E_i,k_i)
\label{eq0110}
\eeq
and $t=-q^2$.
Summations inside $\delta$-functions go over energies and momenta of  the incoming and outgoing reggeons.

The full Green function is a product of $\Gamma^{(n,m)}$ and $n+m$ propagators, that is the inverse
$\Gamma_R^{1,0,1,0}\equiv \Gamma_R^{(1)}$ for the pomerons and $\Gamma_R^{0,1,0,1}\equiv \Gamma_R^{(2)}$  for the odderons.

Let us start with the case when the Green function does not contain  disconnected parts. Then
\beq
G_R^{(nm)}(E_i,k_i)=\Gamma_R^{(nm)}(E_i,k_i)
\prod_{i=1}^{n_P}\Big(\Gamma_R^{(1)}(E_i,k_i)\Big)^{-1}\prod_{i=1}^{n_O}\Big(\Gamma_R^{(2)}(E_i,k_i)\Big)^{-1},
\label{eqad1}\eeq
where $\Gamma_R$ are connected amputated Green functions considered previously.

Our aim is to use the scaling properties of $G_R$ in the integrand. For simplicity we shall consider the simpler case when $\delta_1=\delta_2=0$.
Putting (\ref{gamrphi}) and (\ref{eq087}) with $\delta_{1,2}=0$ into  (\ref{eqad1}) we find the scaling properties of $G_R^{(nm)}$
\[
G_R^{(nm)}(E_i,k_i)=E_N^{1-n_t}\Big(\frac{E_N}{\alpha'_1}\Big)^{(2-n_t)D/4}\xi^c
\Phi^{(nm)}\Big(-\frac{E_i}{E},\xi^{-z}\frac{{k}_i{k}_j}{E_N}\alpha'_1,g_c\Big)\]
\beq
\prod_{i=1}^{n_P}\Big[\Phi_1\Big(\xi^{-z}\frac{k_i^2}{E_N}{\alpha'_1},g_c\Big)\Big]^{-1}
\prod_{i=1}^{n_O}\Big[\Phi_2\Big(\xi^{-z}\frac{k_i^2}{E_N}{\alpha'_1},g_c\Big)\Big]^{-1},
\label{eqad2}\eeq
where
\[\xi=-E/E_N, \ \ c=1-n_t+\frac{1}{2}\gamma_1 n_P+\frac{1}{2}\gamma_2n_O+z(2-n_t)\frac{D}{4}.\]

To extract the total dependence of $I^{(nm)}(E,t)$ we make a change of integration variables
\[E_i=E\zeta_i,\ \ {k}_i=\xi^{z/2}{x}_i.\]
Then we get
\beq
I^{(nm)}(E,t)=E^{-1+a}F^{(nm)}(t\xi^{-z}),
\label{eqad3}
\eeq
where
\beq a=\frac{1}{2}\gamma_1n_P+\frac{1}{2}\gamma_2n_O+\frac{1}{4}zD(n_t-2)
\label{eqad4}\eeq
and some functions  $F(t\xi^{-z)})$, which
are determined  by functions $\Phi$ including also factors from the definition of $\xi$ in terms of $E$.

The amplitude is obtained as the inverse Mellin transform. For given $(nm)$
\[
{\cal A}^{(nm)}(s,t)=A^{n_1,n_2}B^{m_1,m_2}\frac{s}{2\pi i}\int_{\sigma-i\infty}^{\sigma+i\infty}dEe^{-Ey}  I^{(n,m)}(E,t)\]\beq
=\frac{s}{2\pi i}\int_{\sigma-i\infty}^{\sigma+i\infty}\frac{dE}{E}e^{-Ey}E^a\tilde{F}(t(-E)^{-z}E_N^z),
\label{eqad5}\eeq
where $y=\ln s$ and we $\tilde{F}$ includes the impact factors $A$ and $B$.
Changing integration variables $Ey=\epsilon$ we get

\[{\cal A}^{(nm)}(s,t)=sy^{-a}\frac{1}{2\pi i}\int_{\sigma'-i\infty}^{\sigma'+i\infty}
\frac{d\epsilon}{\epsilon}\epsilon^ae^{-\epsilon}\tilde {F}\Big(t y^z \Big(\frac{-\epsilon}{E_N}\Big)^{-z}\Big).\]
Denoting the result of integration over $\epsilon$ as $\Psi(ty^z)$ we find our final result
\beq
{\cal A}^{(nm)}(s,t)=sy^{p(n_p,n_o)}\Psi(ty^z),
\label{eqad6}\eeq
where the power $p=-a$ is
\beq
p(n_P,n_O)=z\frac{D}{2}-n_P\Big(\frac{1}{2}\gamma_1+z\frac{D}{4}\Big)-n_O\Big(\frac{1}{2}\gamma_2+z\frac{D}{4}\Big).
\label{eqad7}
\eeq

We take $D=2$. Then for the simplest exchanges we get for
one pomeron exchange
$p(2,0)=-\gamma_1$
and for one odderon exchange
$p(0,2)=-\gamma_2$.
Exchange by one more pomeron gives the change of power
\[\Delta_P=p(n_P+1,n_O)-p(n_P,n_O)=-\frac{1}{2}(\gamma_1+z).\]
Exchange by two more odderons gives the change of power
\[\Delta_O=p(n_P,n_O+2)-p(n_P,n_O)=-(\gamma_2+z).\]

One can show that these results do not change if  the Green function contains disconnected parts.

The further  study of the asymptotical behaviour (\ref{eqad6}) depends on the numerical values of the anomalous dimensions
at different fixed points.

%

\subsection{At fixed point and  with $D=2$}
Values of $\gamma_1$, $\gamma_2$, $\tau_1$ and $z$ for the five found real fixed points
can be easily calculated using data from the left Table in Section~4.2.
From these values at $D=2$ we find for different points.

\[g_c^{(0)}:\ \ \ p(2,0)=-1,\ \ p(0,2)=0, \
\Delta_p=-\frac{1}{2},\ \ \Delta_O=0.\]

\[g_c^{(1)}:\ \ \ p(2,0)=\frac{1}{6},\ \ p(0,2)=0,\ \
\Delta_p=-\frac{11}{24},\ \ \Delta_O=-\frac{13}{12}.\]

\[g_c^{(2)}:\ \ \ p(2,0)=\frac{1}{6},\ \ p(0,2)=\frac{2}{11.3},\ \
\Delta_p=-\frac{11}{24},\ \ \Delta_O=-\Big(\frac{13}{12}-\frac{2}{11.3}\Big).\]

\[g_c^{(3)}:\ \ \ p(2,0)=\frac{1}{6},\ \ p(0,2)=\frac{1}{12},\ \
\Delta_p=-\frac{11}{24},\ \ \Delta_O=-1.\]

\[g_c^{(4)}:\ \ \ p(2,0)=-1,\ \ p(0,2)=0,\ \
\Delta_p=-1,\ \ \Delta_O=-1.\]

Inspecting these results we find the following.

$\bullet$ All $\Delta_P$ are negative. So the leading contribution comes from
the minimal number of exchanged pomerons.

$\bullet$ For all fixed points except $g_c^{(0)}$ also $\Delta_O$ is negative,
so that the leading contribution comes from the minimal number of exchanged odderons.
At the fixed point $g_c^{(0)}$ we find $\Delta_O=0$ and the asymptotic is the same for any number of exchanged odderons.

$\bullet$ At $g_c^{(1,2,3)}$ the cross-sections due to the single pomeron exchange grow as $y^{1/6}$.
At $g_c^{(0,4)}$ the cross-sections fall as $1/y$.

$\bullet$ At $g_c^{(0,1,4)}$ the cross-sections due to the single odderon exchange are constant.
At $g_c^{(3)}$ it rises as $y^{1/12}$, however, not so fast as the one-pomeron exchange ($\sim y^{1/6}$).
Notably at $g_c^{(2)}$ the odderon cross-section is rising  as $y^{2/11.3}$, slightly faster than the pomeron exchange.
However, one should have in mind that $g_c^{(2)}$ is a strongly repulsive fixed point hardly reachable in evolution.

$\bullet$ So finally, in all important cases when the single pomeron contribution grows it dominates over all multireggeon
contributions, as in absence of odderons, which result was found in ~\cite{abarb1}.

Taking into account that the only totally attractive fixed point is $g_c^{(3)}$ we conclude
from our study that
most probably the leading contribution will be the single pomeron exchange and the subdominant one the single odderon exchange
\beq
{\cal A}(s,t)=sy^{1/6}\Psi_{20}(ty^{13/12})+sy^{1/12}\Psi_{02}(ty^{13/12})
\label{eqad10}\eeq
with the cross-section of the form
\beq
\sigma^{tot}=y^{1/6}A+y^{1/12}B+{\cal O}(y^{-1/3}).
\label{eqad11}\eeq

Note that this cross-section describes  scattering of effectively pointlike particles (protons), as illustrated in Fig. 6, $a$ and $b$.
\begin{figure}
\begin{center}
\epsfig{file=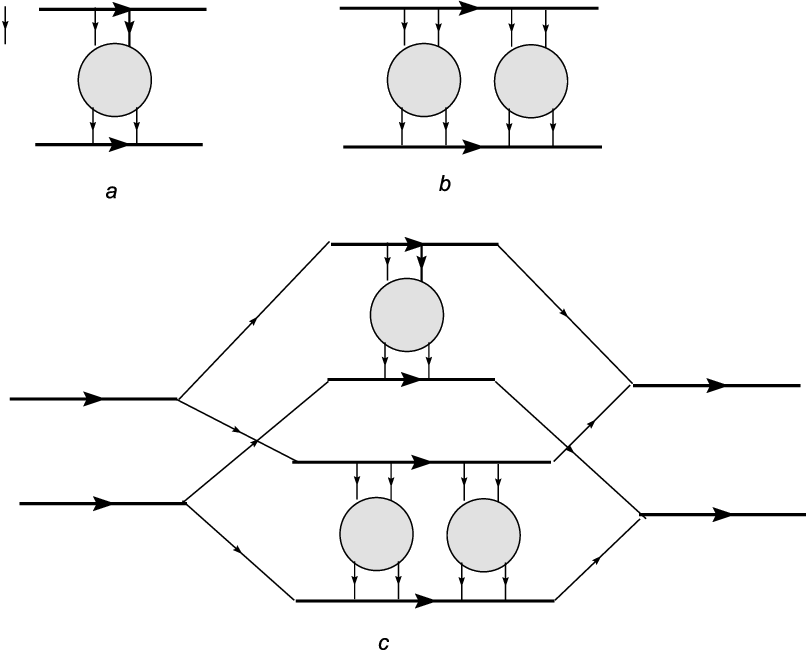, width=8 cm}
\caption{Elasic scattering for effective pointlike particles (protons), $a$ and $b$, and composite particles (nuclei), $c$.}
\end{center}
\label{fig6}
\end{figure}
For nuclei the amplitude involves disconnected parts describing collisions of their components, Fig. 6, $c$.
For heavy nuclei these elementary amplitudes exponentiate and our expressions refer to what is termed  eikonal function.

\section{Conclusions}
We have constructed a Regge-Gribov model of interacting pomerons and odderons.
For the pomeron the overall influence of interactions with odderons has been found weak and
of no fundamental character. In particular, in the zero-dimensional transverse space the possibility for analytic continuation
to the intercept greater than unity is intact. The lowest energy remains positive exponentially small in $1/\lambda$.
The interaction with odderons makes this energy even smaller by 30\%.
With two dimensions the pomeron propagator  feels odderons only at certain fixed points which, however, are not fully attractive.
At the only fully attractive fixed point $g_c^{(3)}$  influence of the odderon on pomeron disappears (since the coupling constant
for pomeron-odderon vanishes at this fixed point). As a function of mass parameters $\delta_{1,2}$ at all fixed points
the Green functions have a singularity
of the type $\delta^p$ where $p$ is a rational number, so that continuation to $\delta<0$  meets with trouble and indicates a phase transition
to unphysical phases. So in this respect the model with odderons exists only for $\delta_{1,2}\geq 0$ just as without them.
The elastic cross-section (and the total one) are dominated by the single pomeron and odderon exchanges and so lead to the total
cross-section
growing as $(\ln s)^{1/6}$ with the subdominant odderon contribution which also  grows as $(\ln s)^{1/12}$

In the renormalization group approach five real fixed points have been found.
Of all of them only  one $g_c^{(3)}$ is purely attractive.  Numerical probes to study trajectories starting from homogeneously distributed points
in the 4-dimensional domain of coupling constants have shown that
in 35\% of all cases it was impossible to follow
the trajectories far away from  fixed points and so they went to large values of coupling constants beyond our lowest order approximation.
In the rest 65\% cases the distribution  of the arrival at specific fixed points was found to be in percentage
\[g_c^{(0)}:g_c^{(1)}:g_c^{(2)}:g_c^{(3)}:g_c^{(4)} \ =\ 0:0.33:0:92.6:7.02.\]
So in fact in the vast majority of cases the trajectories arrive at  $g_c^{(3)}$ confirming its exceptional quality.
It would be desirable to advance calculations to study  double loops. This may shed light on the cases when the trajectories go away from
the fixed points found with the single loop approximation. This formidable task is postponed for future studies.

\section{Conflicts of interest}

The authors declare that they have no conflicts of interest.

\section{Acknowledgements}

The authors are grateful to N.V. Antonov, N.M. Gulitskiy and P.I. Kakin
for very useful discussions.
We also thanks the organizers of the VIII International Conference
''Models of Quantum Field Theory'' (2025).


\end{document}